\documentclass{article}
\usepackage{spconf,amsmath,graphicx}
\usepackage{url}
\usepackage{tabularx}
\usepackage{multirow}
\usepackage{xcolor,colortbl}
\usepackage{xspace}
\usepackage{comment}
\usepackage{subcaption}
\usepackage{tipa}
\usepackage{textcomp}
 
 \usepackage[normalem]{ulem}
 \usepackage{booktabs}

\definecolor{neutralgreen}{HTML}{1B9AAA}
\definecolor{positivegreen}{HTML}{06D6A0}
\definecolor{negative}{HTML}{EF476F}

\arrayrulecolor{positivegreen}

\usepackage[colorlinks=true,allcolors=negative]{hyperref}

\newcommand{\sota}{state-of-the-art\xspace}

\title{Modelling low-resource accents without accent-specific TTS frontend}
\name{Georgi Tinchev$^*$$^1$, Marta Czarnowska$^*$$^1$, Kamil Deja$^2$, Kayoko Yanagisawa$^1$, Marius Cotescu$^1$}
\address{$^1$Amazon Research, United Kingdom $^2$Warsaw University of Technology, Poland
}

\begin{document}
%
\maketitle

\def\thefootnote{*}\footnotetext{Indicates equal contribution. $^2$Work done while at Amazon.}\def\thefootnote{\arabic{footnote}}
\footnotetext[3]{https://bit.ly/3V52ZrF}
\begin{abstract}This work focuses on modelling a speaker's accent that does not have a dedicated text-to-speech (TTS) frontend, including a grapheme-to-phoneme (G2P) module. Prior work on modelling accents assumes a phonetic transcription is available for the target accent, which might not be the case for low-resource, regional accents. In our work, we propose an approach whereby we first augment the target accent data to sound like the donor voice via voice conversion, then train a multi-speaker multi-accent TTS model on the combination of recordings and synthetic data, to generate the donor's voice speaking in the target accent. Throughout the procedure, we use a TTS frontend developed for the same language but a different accent. We show qualitative and quantitative analysis where the proposed strategy achieves \sota results compared to other generative models. Our work demonstrates that low resource accents can be modelled with relatively little data and without developing an accent-specific TTS frontend. Audio samples of our model converting to multiple accents are available on our web page\footnotemark[3].
\end{abstract}

\vspace{-.5em}
\section{Introduction}
\label{sec:intro}
\vspace{-.35em}

Recent advancements in text-to-speech (TTS) have been made to explore multilingual training of the acoustic models~\cite{multilingual_tacotron2,polyglot2022phonemes}. They usually rely on huge amounts of multi-lingual and multi-speaker data and can convert between individual accents by conditioning specifically on the speaker and accent information. While attempts have been made to condition the model directly on character input~\cite{tacotron2}, many of the approaches still rely on phoneme input since it (a) avoids pronunciation errors introduced by the model mislearning how to pronounce certain words and makes the task for the neural network easier~\cite{pmlr-v80-skerry-ryan18a}; and (b) allows for phonetic control of the synthesized speech~\cite{perquin2021}. Phoneme sequences can be generated by passing the text input through a TTS frontend. The latter consists of several modules including text normalisation and grapheme-to-phoneme (G2P) conversion, and is typically language- or accent-dependent. It is expensive to develop a dedicated frontend due to the need for expert language knowledge~\cite{kenyan_english_dialect}. 

In this paper we propose a TTS system that is capable of modelling a target accent with a selected donor voice without the need for a dedicated frontend that can produce phonetic transcriptions matching the target accent. Our model produces~\sota results in terms of naturalness and accent similarity. The model generalizes to learn differences in phonological contrasts between accents even when trained with phoneme sequences produced by a frontend not specifically developed for the accent. In summary, our contributions are as follows: 1) We propose the first approach, to our knowledge, for accent modelling without explicitly conditioning the model on phonemes generated with an accent-specific G2P for the target accent. Our method demonstrates \sota results compared with existing multi-lingual multi-accent baselines, modelled by normalizing flows, VAEs, or diffusion models, 2) we can reliably model the accent with augmented data regardless of the chosen TTS frontend and 3) we demonstrate that modelling the accent via synthetic data can be done with just 2000 utterances of the target accent.

\vspace{-1.25em}
\section{Related Works}
\vspace{-.75em}

This section summarises relevant literature for the task of accent modelling. The review is split in two sections - research in grapheme-to-phoneme (G2P) and accent conversion.

TTS models usually use phoneme sequences as inputs to model speech, since directly modelling from character inputs adds an extra challenge for the network to learn the relationship between characters and pronunciation. Work on G2P traditionally involved rule-based methods~\cite{cmu_dict,nettalk,kingsbury1997callhome}, then evolved to machine learning based approaches using n-grams~\cite{novak_minematsu_hirose_2016} or neural networks~\cite{yao_zweig_2015,rao2015}. Recently, there have been efforts to translate these methods to low-resource languages by using distance measures and linguistic expertise~\cite{deri2016grapheme}. The main focus of our work is to develop a TTS system for an accent for which we do not have a dedicated frontend, as developing a new frontend requires linguistic expertise and is not scalable.

Historically accent conversion (AC) has been achieved in two ways - combining spectral features via voice morphing~\cite{kayoko_accent_morphing,felps2009foreign,foreign_ac_voice_morph} and voice conversion (VC) adaptation using spectral and phoneme frame matching~\cite{vc_ac,e2e_ac,accent_disentanglement_vc}. There is also a line of work that relies on external features with frame-matching after the VC step~\cite{ac_ppg,zhao2019foreign,ppg_segmental_ac,ding2022accentron,sabr, sabr_ac}. The main shortcomings of voice morphing methods is their requirement for parallel data. For frame-matching it is hard to convert the duration and speaking rate of the target voice. In contrast to existing VC-based approaches for AC, we do not require an ASR model to disentangle the accent information. Instead we leverage VC as a first step to generate data of the \textit{donor voice} in the \textit{target accent}, since it preserves the accent~\cite{vc_ac}. 

Similarly to us, in the domain of speech recognition with foreign accent, other works observe that data augmentation through voice and accent conversion positively impacts the model's performance~\cite{fukuda2018data,sjtu}. Most similar to us,~\cite{google_accent_transfer} also leverage data augmentation for accent transfer. Crucially, our model does not require text to phoneme transcriptions for the \textit{target accent} and we are able to model the accent by using less data - just 27k utterances vs 425k utterances in~\cite{google_accent_transfer}.

\vspace{-1.25em}
\section{Methodology}
\vspace{-.75em}

The goal of our method is to train a TTS model for a new low-resource accent (from here on referred to as the \textit{target accent}) without an accent-specific frontend to generate phonetic transcriptions with. To achieve this we use a multi-speaker multi-accent dataset. We select a \textit{donor} speaker from one of the non-target accent speakers. We want the synthesized speech to sound like the \textit{donor} speaker but with the \textit{target accent}. We require G2P model for one of the accents from the dataset, which we use to extract the phoneme sequence for all the utterances. Crucially, we do not require accent-specific G2P for any of the other accents, including the \textit{target accent}. In this section we will explain the model and how we generate augmented data for the \textit{donor} voice in the \textit{target accent}.

\vspace{-1em}
\subsection{Architecture}

We divide our approach into two major steps - Voice Conversion (VC) and Text-To-Speech (TTS). We use VC to generate synthetic data of our \textit{donor} voice speaking in the \textit{target accent}. We then use this augmented data together with the original recordings to train a multi-speaker multi-accent TTS system. Below we briefly outline each of the two models.

\vspace{-1em}
\subsubsection{Voice conversion}

To augment the training data for the TTS model, we use a flow-based voice conversion model~\cite{flow_tts,spring_vc}. During training the model encodes the input mel-spectrogram into a latent vector $z$, using phoneme sequence, speaker embeddings, f0, and binary voice/unvoiced flag as conditioning. The f0 is normalized per utterance to disentangle speaker information from the utterance prosody. We use pre-trained speaker embeddings based on~\cite{speaker_embeddings}. The model is optimized with a negative log-likelihood loss. During inference we encode the mel-spectrogram of the \textit{target accent} speaker into the latent $z$, using corresponding conditioning, and then decode it back to a mel-spectrogram, changing the speaker embedding to that of the \textit{donor} speaker. We first train the VC model on the whole dataset. Using this model we convert the speaker from the \textit{target accent} to sound like the \textit{donor} voice. This way we are able to change the speaker identity but preserve the rest of the speaking characteristics, namely accent, style, and duration, in line with~\cite{vc_ac}. We then train TTS acoustic and duration models on the original dataset and synthetic data generated by the VC model, as described in the following section.

\begin{figure}[t]
\centering
    \includegraphics[width=0.8\linewidth]{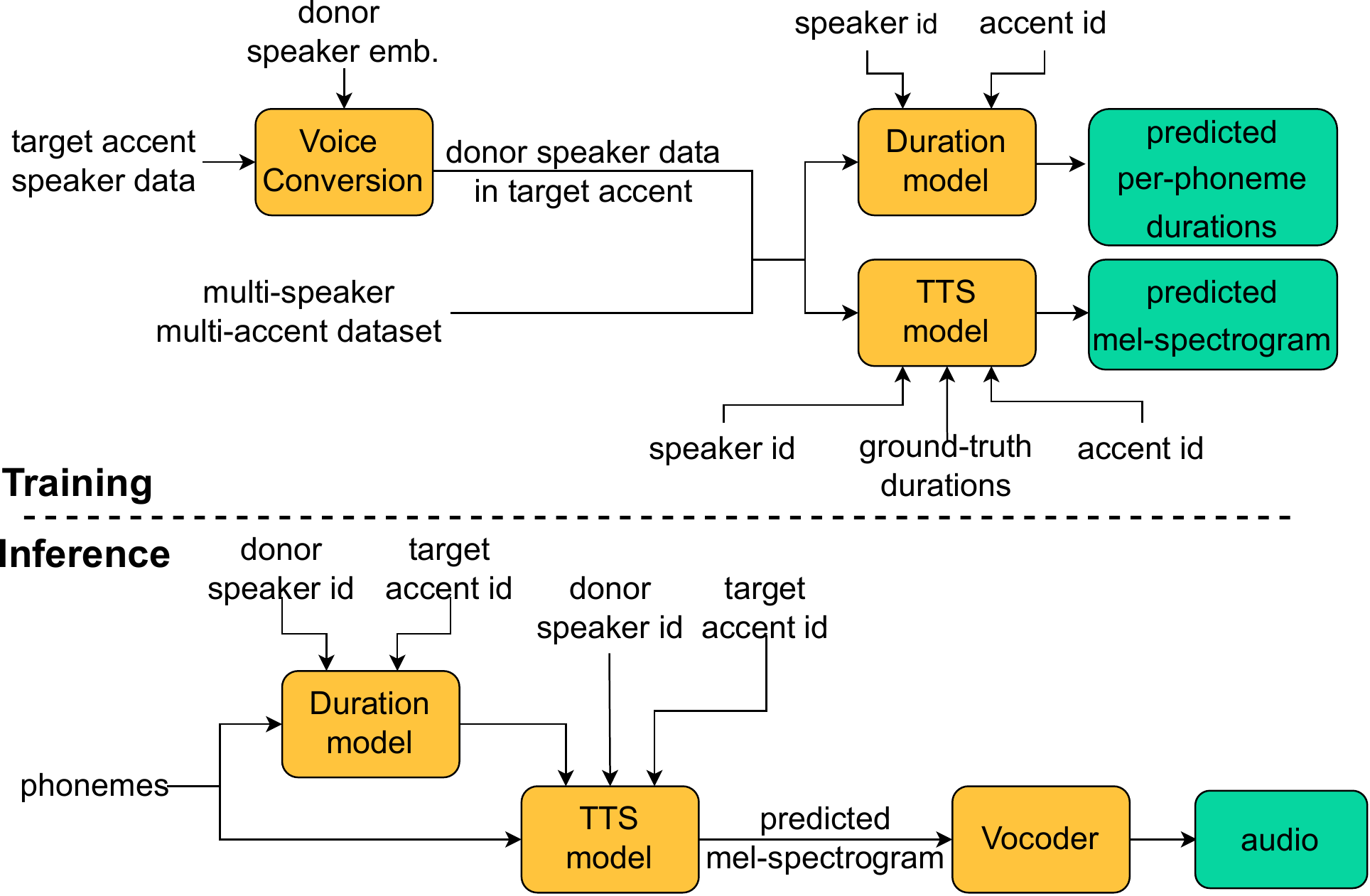}
    \caption{TTS model training and inference procedure}
    \label{fig:train_and_inf_flow}
\vspace{-2em}
\end{figure}

\begin{table*}[t]
\caption{MUSHRA comparison to the \sota. For accent similarity we used 25 test cases from 4 en-IE speakers as reference. We used two upper anchors: i) different recording of the reference speaker ii) recording of a different en-IE speaker.}
\label{table:exp2_sota_baselines}
\vspace{-2em}
\begin{center}
\resizebox{\textwidth}{!}{
    \begin{tabular}{lcccccc}
    \toprule
    \multirow{2}{*}{System} & \multirow{2}{*}{Naturalness} & Accent similarity & Accent similarity & Accent similarity & Accent similarity & Accent similarity \\
    & & all speakers & speaker 1 & speaker 2 & speaker 3 & speaker 4 \\
    \midrule
    Upper anchor & 74.42 & 82.24 & 72.56 & 89.33 & 84.79 & 79.24 \\
    
    Upper anchor different speaker & N/A & 52.55 & 43.40 & 49.42 & 57.10 & 54.89\\
    
    Proposed method & \textbf{64.61} & \textbf{50.73} & \textbf{61.40} & \textbf{40.81} & \textbf{45.16} & \textbf{50.44} \\
    
    Grad-TTS & 63.23 & \textbf{50.68} & 59.29 & \textbf{40.77} & \textbf{46.32} & \textbf{50.71} \\
    
    Flow-TTS & 56.46 & \textbf{50.32} & 59.09 & \textbf{40.28} & \textbf{45.19} & \textbf{51.21} \\
    \bottomrule
    \end{tabular}
}
\vspace{-3em}
\end{center}
\end{table*}

\vspace{-1em}
\subsubsection{Text-to-speech}
Our text-to-speech (TTS) architecture is a sequence-to-sequence model with explicit duration~\cite{tacotron2,edp}. It has two separate models: the acoustic model and the duration model.

The acoustic model consists of three components: an encoder, an upsampling layer, and a decoder. The encoder architecture is based on Tacotron2~\cite{tacotron2} and the decoder is composed of residual gated convolution layers and an LSTM layer. We use a Flow-VAE reference encoder~\cite{polyglot2022phonemes}. We train this architecture with 2000 utterances of the \textit{donor speaker} in the \textit{target accent}, generated with VC. However, we empirically found out that training on a multi-speaker multi-accent dataset improves the naturalness (Section~\ref{sec:exp_ablations}). We therefore modify the architecture to include speaker and accent conditioning. We condition both the encoder and decoder on speaker and accent embeddings, trained together with the model. During training, ground-truth phoneme durations are used to upsample the encoded phonemes sequence, but during inference we use the durations predicted by the duration model. The acoustic model is optimized with an L2 loss between ground truth and predicted mel-spectrogram and KL divergence for the VAE.

The duration model uses the same encoder architecture as the acoustic model, followed by a projection layer. The model is also conditioned on speaker and accent information. It is optimized with an L2 loss between ground-truth and predicted phoneme durations. During inference we predict the duration of the phonemes conditioned on the \textit{target accent} and the \textit{donor} speaker. We then run inference with the acoustic model, providing the same speaker and accent conditioning, predicted durations, and VAE centroid computed from the synthetic data generated by the VC model. We vocode predicted mel-spectrograms using a universal vocoder~\cite{universal_vocoder}.

\begin{table}[t!]
\caption{MUSHRA comparison to the \sota.}
\label{table:exp1_polyglot_basline}
\vspace{-2em}
\begin{center}
    \resizebox{\linewidth}{!}{%
        \begin{tabular}{lcc}
        \toprule
        System & Naturalness & Accent similarity \\
        \midrule
        Upper anchor different speaker & 90.46 & 65.85 \\
        
        Proposed method & \textbf{72.33} & \textbf{74.14} \\
        
        Polyglot & 47.24 & 49.29 \\
        
        Lower anchor (en-GB) & N/A & 12.17  \\
        \bottomrule
        \end{tabular}
    }
\vspace{-3em}
\end{center}
\end{table}

\vspace{-1em}
\subsection{Dataset}

In our experiments we chose to model Irish English (en-IE) accent using a British English (en-GB) \textit{donor speaker}. We use 25k utterances for the \textit{donor speaker} and between 0.5k - 4.5k utterances for the rest of the speakers from 6 supporting accents - British (en-GB), American (en-US), Australian (en-AU), Indian (en-IN), Welsh (en-GB-WLS), and Canadian English (en-CA)~\cite{polyglot2022corpus}. For en-IE, which is our \textit{target accent}, we have the same 2000 utterances recorded by 12 speakers. Having this type of parallel data is not a requirement for either the VC or TTS models. In Section~\ref{sec:exp_ablations} we demonstrate that training on just 2000 utterances by a single speaker is enough for our approach to model the accent. We extracted phonetic transcriptions using the en-GB frontend, which is the accent of the \textit{donor speaker}. We use a unified representation for the phonemes - we map the same phonemes across different accents to the same input tokens~\cite{polyglot2022phonemes}.

\vspace{-1.25em}
\section{Experiments}
\vspace{-.75em}

In our work we have chosen to model en-IE accent with an en-GB \textit{donor speaker}. In this section we demonstrate how our method achieves \sota results compared to other TTS models. We also present qualitative and quantitative evaluations of the proposed model using two different G2P models to generate phonetic transcriptions: en-GB and en-US. Finally, through ablation studies, we will demonstrate that modelling accents can be done with low-resource data.

\vspace{-1em}
\subsection{Evaluation method}

We used a MUltiple Stimuli with Hidden Reference and Anchor (MUSHRA) test for comparing our proposed method to \sota baselines~\cite{mushra}. Each evaluation consisted of 100 unique testcases, not seen during training of the models. The testcases were evaluated by 24 native Irish speakers who rated each system on a scale between 0 and 100. We evaluated naturalness and accent similarity of the samples. For naturalness testers were asked to judge if the samples sound like a human speaker and for accent similarity how similar are the samples to a provided reference sample. Both the reference and the hidden upper anchor were the recordings of en-IE speakers. To ensure results are statistically significant, a paired t-test with Holm-Bonferroni correction was performed ($p \leq 0.05$). Bold results indicate the best performing systems in each column, up to statistically significant differences.

\begin{table}[t!]
\caption{Ablation study for different data configurations.}
\label{table:exp4_data_ablation}
\vspace{-2em}
\begin{center}
    \resizebox{\linewidth}{!}{%
        \begin{tabular}{lcc}
        \toprule
        System & Naturalness & Accent similarity \\
        \midrule
        VC multi-speaker multi-accent & 78.58 & 60.85 \\
        
        TTS multi-speaker multi-accent & \textbf{65.07} & \textbf{58.12} \\
        
        TTS single speaker IE finetuned & 59.07 & \textbf{57.32} \\
        
        Lower anchor (en-GB) & N/A & 11.50  \\
        \bottomrule
        \end{tabular}
    }
\vspace{-3em}
\end{center}
\end{table}

\vspace{-1em}
\subsection{Comparison to the \sota}
\label{ref:exp_polyglot_representations}

We compared our method to Polyglot - a multi-speaker multi-accent sequence-to-sequence attention-based model~\cite{polyglot2022corpus}. It achieves \sota results using phonetic transcriptions generated by accent specific G2P. We trained both our approach and the baseline using phonetic features extracted with an en-GB G2P model. Table~\ref{table:exp1_polyglot_basline} shows our method achieving significantly better results in naturalness and accent similarity, indicating that accent and speaker conditioning alone are not sufficient for accent transfer without accent specific phonetic transcriptions. In the accent similarity evaluation we used a different en-IE speaker for the reference and the upper anchor, both had the same variety of en-IE accent. The evaluators were good at distinguishing this subtle difference and thus rated our model better than the upper anchor. Therefore, in following experiments we introduced a second upper anchor with different recordings of the reference speaker.

We also compared our method to two additional baseline models, trained on the same dataset with phonemes for each accent extracted with en-GB G2P: 1) Grad-TTS~\cite{grad_tts} with explicit speaker and accent conditioning and 2) Flow-TTS-based accent conversion model~\cite{spring_ac,flow_tts}. The results are presented in Table~\ref{table:exp2_sota_baselines}. For naturalness our method achieves significantly better results than both baselines.
As Grad-TTS and Flow-TTS baselines model a generic en-IE accent, instead of speaker-specific en-IE accent, we decided to use 4 different en-IE speakers as the reference speaker. Our method is significantly better when the reference is the same en-IE speaker (we used speaker 1's accent for data augmentation), but there is no significant difference between the models for different reference speakers (speaker 2, 3, 4). This confirms that by using VC samples our proposed approach is better tailored towards a particular speaker's accent.

\begin{figure}[t]
\centering
\begin{subfigure}{.4\columnwidth}
  \centering
  \includegraphics[width=0.9\columnwidth]{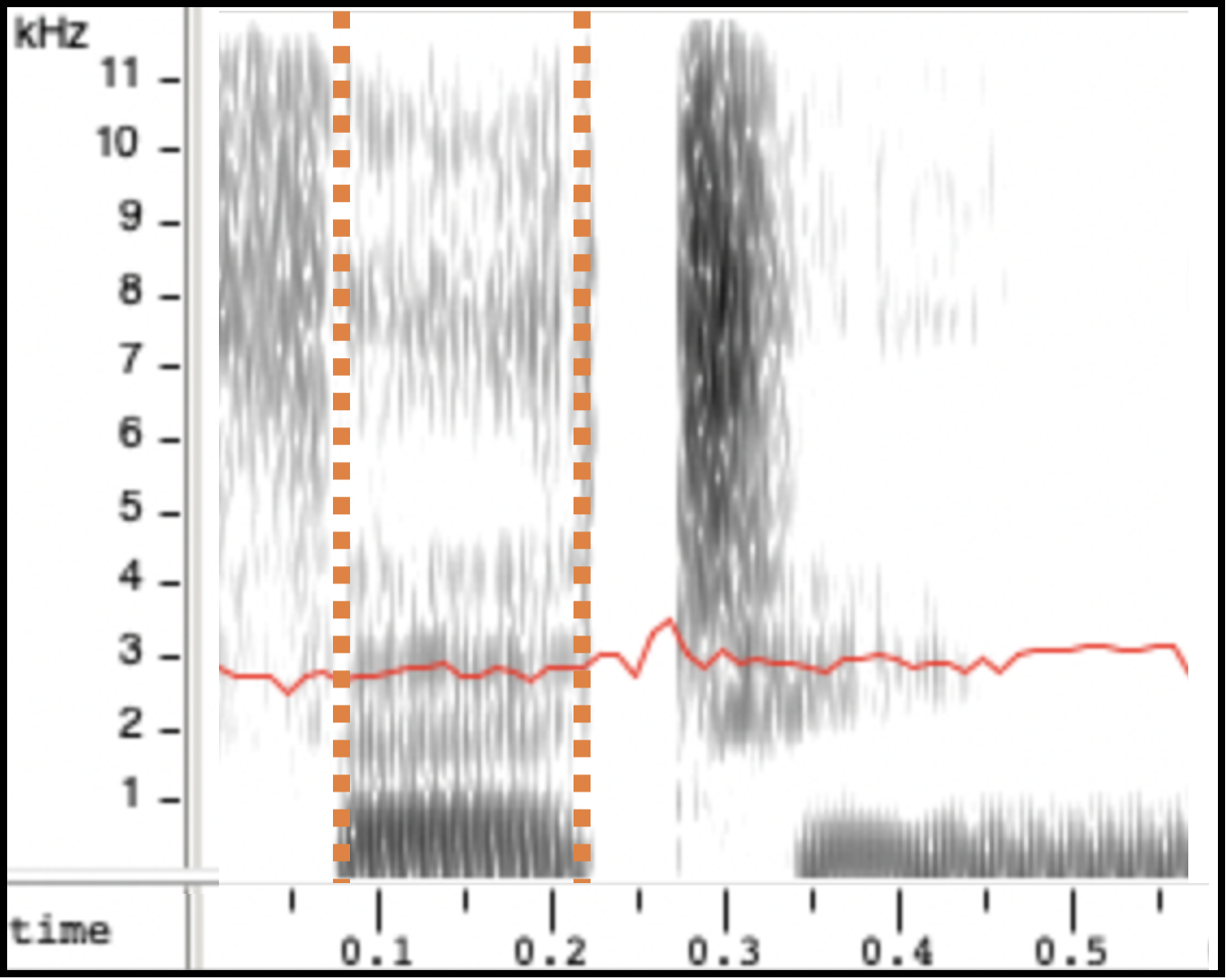} 
  \caption{en-GB synthesis}
  \label{fig:f3_spec1}
\end{subfigure}
\begin{subfigure}{.4\columnwidth}
  \centering
  \includegraphics[width=0.9\columnwidth]{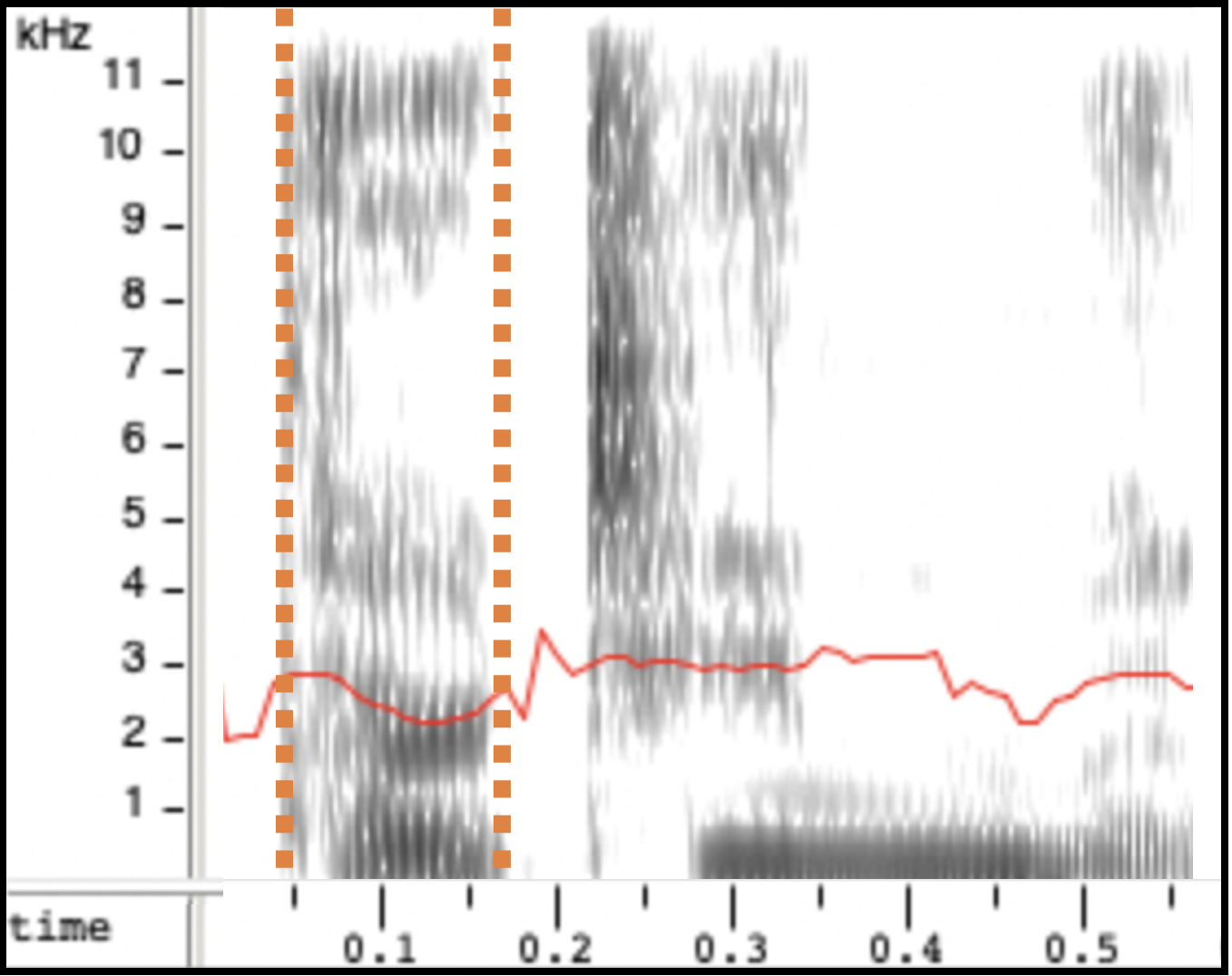}
  \caption{en-IE ground truth}
  \label{fig:f3_spec2}
\end{subfigure}
\newline
\begin{subfigure}{.4\columnwidth}
  \centering
  \includegraphics[width=0.9\columnwidth]{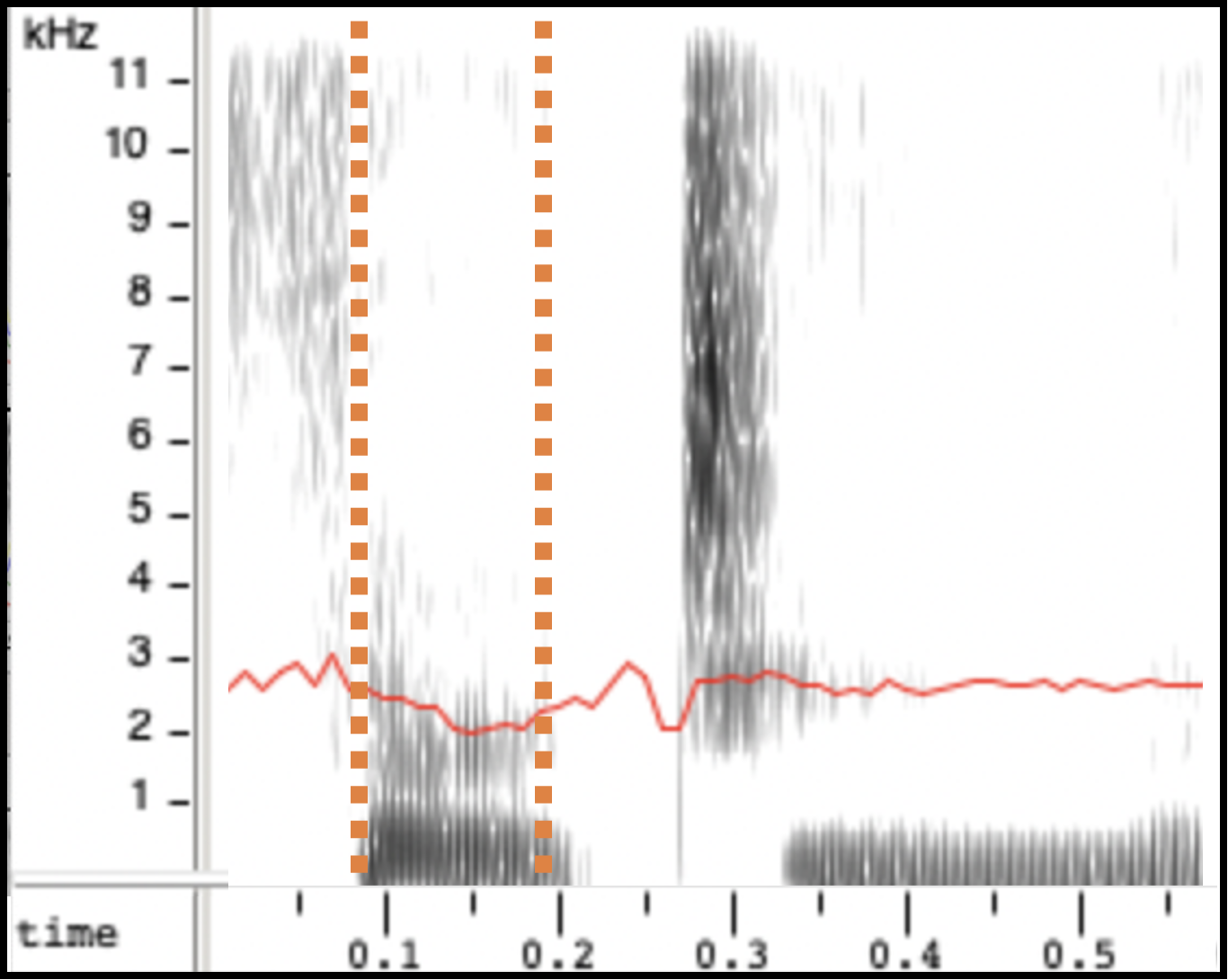}
  \caption{en-IE with en-GB G2P}
  \label{fig:f3_spec3}
\end{subfigure}
\begin{subfigure}{.4\columnwidth}
  \centering
  \includegraphics[width=0.9\columnwidth]{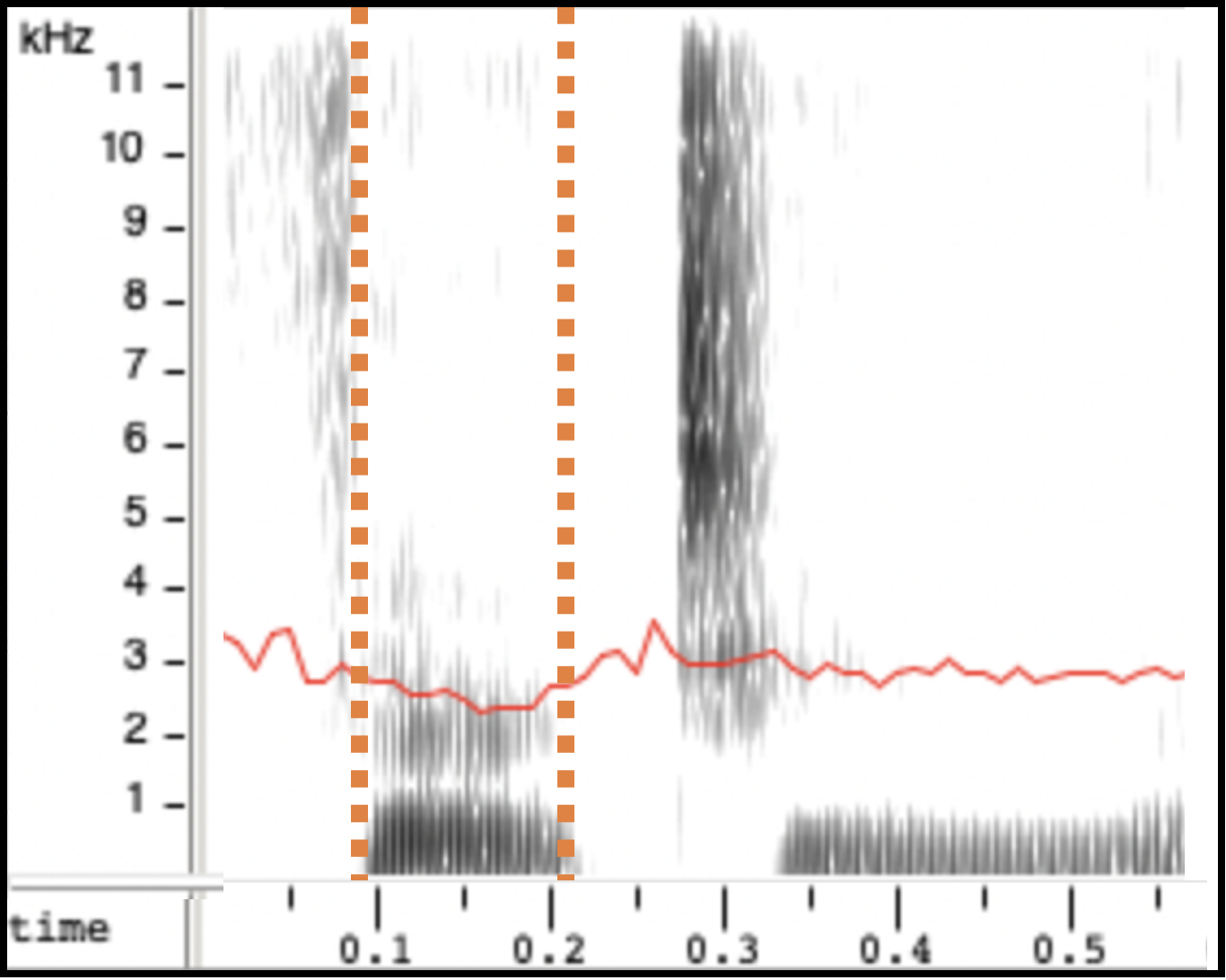}
  \caption{en-IE with en-US G2P}
  \label{fig:f3_spec4}
\end{subfigure}
\newline
\begin{subfigure}{.725\columnwidth}
  \centering
  \includegraphics[width=\textwidth]{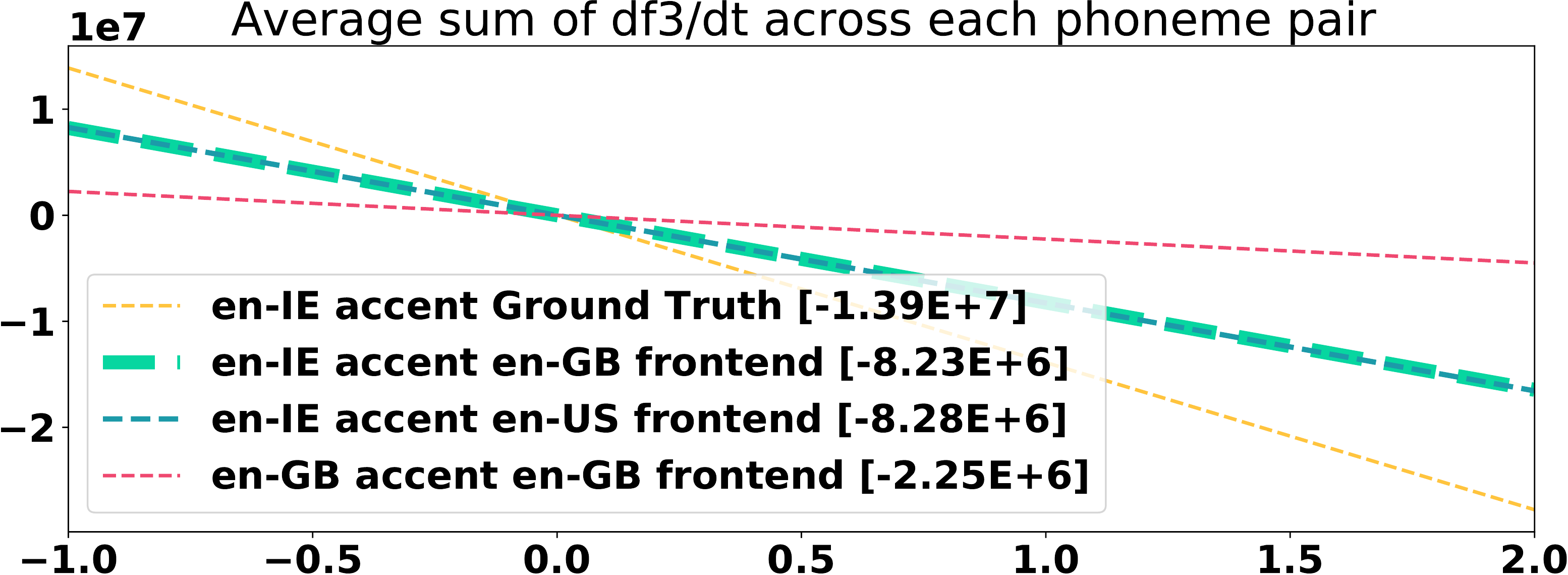}
  \caption{Average sum of F3 derivatives }
  \label{fig:f3_slopes}
\end{subfigure}
\vspace{-0.5em}
\caption{(a-d) F3 contour (in red) showing the lowering of the formant in the \textipa{/\textrhookrevepsilon/} or \textipa{/3:/} vowel in ``thirteen'', indicated between orange dashed lines. (e) Illustration of the average slope of the postvocalic /r/ contexts.}
\label{fig:f3_spec}
\vspace{-1.5em}
\end{figure}

\vspace{-1em}
\subsection{Analysis of Rhoticity}
\label{sec:exp_frontend}

Next, we show how we can reproduce features of the \textit{target accent} when we use a frontend for an accent that is missing that feature. In all our experiments above, we presented the results for the model trained with en-GB G2P (``en-IE with en-GB G2P''), the accent of the \textit{donor voice}. This model is capable of generating samples with an accent closer to en-IE than en-GB (lower anchor), as shown in Table~\ref{table:exp1_polyglot_basline}. en-IE differs from en-GB in one major aspect, rhoticity, which means that the /r/ is pronounced in postvocalic contexts when not followed by another vowel (e.g. \textipa{/kA:r.pA:rk/} for ``car park''). en-GB, on the other hand, is non-rhotic (\textipa{/kA:.pA:k/}). Despite this difference, our model trained with en-GB G2P was able to reproduce the rhoticity in the synthesised en-IE samples.

To quantify this, we analysed the third formant (F3) for the segments where the rhoticity contrast is found, as lowering of F3 is acoustically correlated to the phoneme /r/~\cite{r_f3}. For comparison, we trained an additional model where the phonemes were extracted with the en-US G2P, which is also a rhotic accent (``en-IE with en-US G2P''). To identify the regions where the contrast occurs, we extracted phonemes for our test set with both en-GB and en-US frontends and aligned the phoneme sequences using Dynamic Time Warping (DTW)~\cite{dtw} with a cost function based on phoneme similarity. We use the alignment to identify contexts for which the /r/ is present in rhotic accents but not in non-rhotic accents. For example, in the phrase "car park", en-GB \textipa{/A:/} is aligned with en-US \textipa{/A:r/}. We use Kaldi external aligner~\cite{kaldi} to find each phoneme position in the audio file and LPC analysis to extract the F3 for those contexts and compute its slope.

\begin{figure}[t]
\centering
\includegraphics[width=\linewidth]{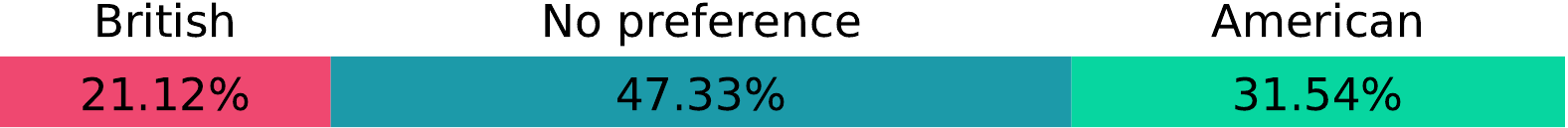}
\vspace{-2em}
\caption{Preference test results for samples from model trained with en-GB and en-US frontends.}
\label{fig:engb_vs_enus_pref}
\vspace{-1.5em}
\end{figure}

Fig.~\ref{fig:f3_spec} shows how monolingual en-GB (non-rhotic) model sample does not show a lowering of F3 (\ref{fig:f3_spec1}) whereas the ground truth en-IE (rhotic) recording does (\ref{fig:f3_spec2}). en-IE samples generated from models trained with our approach show the lowering of F3, regardless of whether the model was trained with en-GB G2P (\ref{fig:f3_spec3}) or en-US G2P (\ref{fig:f3_spec4}). This is confirmed in Fig.~\ref{fig:f3_slopes} which shows the average gradient of the F3 slope across 134 contexts with the rhotic contrast. There is no statistically significant difference between the gradient for ``en-IE with en-GB G2P'' and ``en-IE with en-US G2P''.

We subsequently compared ``en-IE with en-GB G2P'' and ``en-IE with en-US G2P'' in a preference test with 24 listeners each listening to 100 pairs of test cases. Fig.~\ref{fig:engb_vs_enus_pref} shows that there is a significant preference for the model trained with en-US G2P, although the majority had no preference. These results show that we can reproduce features in the \textit{target accent} even if we use the frontend for an accent that is missing that feature, but since accents can vary in many different aspects, we can further improve our model performance by carefully selecting the frontend used for annotation of the data.

\vspace{-1em}
\subsection{Ablation Studies}
\label{sec:exp_ablations}

As a final experiment we present ablation studies of our two models with different data configurations. We trained our TTS model with only the \textit{donor voice} and fine-tuned with the \textit{target accent} synthetic data. Table~\ref{table:exp4_data_ablation} demonstrates that there is no significant difference between the accent similarity of a multi-speaker, multi-accent model and the model trained on the \textit{donor speaker} and fine-tuned with the \textit{target accent} synthetic data. In line with~\cite{polyglot2022corpus}, training in a multi-speaker, multi-accent scenario improves the naturalness. Results from this experiment show that using data from a single target accent speaker is sufficient for the task of modelling the accent.
\vspace{-2em}
\section{Conclusions}
\label{sec:conclusions}
\vspace{-.75em}

In this work we presented an approach for accent modelling based on data augmentation with voice conversion and text-to-speech model trained on the combination of recording and synthetic data. Crucially, our approach does not require transcriptions generated with accent-specific G2P. We show that our model can reliably model features in the \textit{target accent} and achieves \sota performance compared to existing strategies for accent modelling. We show that our approach works in a low resource scenario where we have little data in the \textit{target accent}. We also found that the performance can be improved further by selecting a more suitable accent for the G2P. The strategy for choosing the best accent for G2P is left for future work. We also plan to extend our methodology to accents of language other than English.

\vspace{-1.25em}
\section{Acknowledgements}
\vspace{-.75em}

We thank Andre Canelas, Tom Merritt, Piotr Bilinski, Maria Corkery, and Dan McCarthy for their feedback and insights.

\bibliographystyle{IEEEbib}
{\footnotesize
    \bibliography{main}
}

\end{document}